\documentclass[preprint2]{aastex62}
\usepackage{graphicx}
\usepackage{mathtools}
\usepackage{xcolor}
\usepackage{booktabs}
\usepackage{color,soul}
\usepackage{amsmath}
\usepackage{hyperref}
\usepackage{times}

\begin{document}

\title{The Circumgalactic Medium from the CAMELS Simulations: Forecasting Constraints on Feedback Processes from Future Sunyaev-Zeldovich Observations}

\author[0000-0003-1593-1505]{Emily Moser}
\affiliation{Department of Astronomy, Cornell University, Ithaca, NY 14853, USA}

\author[0000-0001-5846-0411]{Nicholas Battaglia}
\affiliation{Department of Astronomy, Cornell University, Ithaca, NY 14853, USA}

\author[0000-0002-6766-5942]{Daisuke Nagai}
\affiliation{Department of Physics, Yale University, New Haven, CT 06520, USA}

\author[0000-0001-8914-8885]{Erwin Lau}
\affiliation{Center for Astrophysics $\vert$ Harvard \& Smithsonian, Cambridge, MA, 02138, USA}

\author[0000-0002-1948-3562]{Luis Fernando Machado Poletti Valle}
\affiliation{Institute for Particle Physics and Astrophysics, ETH Z\"urich, Wolfgang-Pauli-Strasse 27, CH-8093 Z\"urich, Switzerland}

\author[0000-0002-4816-0455]{Francisco Villaescusa-Navarro}
\affiliation{Center for Computational Astrophysics, Flatiron Institute, 162 5th Avenue, New York, NY, 10010, USA}
\affiliation{Department of Astrophysical Sciences, Princeton University, Peyton Hall, Princeton NJ 08544, USA}

\author[0000-0002-4200-9965]{Stefania Amodeo}
\affiliation{Université de Strasbourg, CNRS, Observatoire astronomique de Strasbourg, UMR 7550, F-67000 Strasbourg, France}

\author[0000-0001-5769-4945]{Daniel Angl\'es-Alc\'azar}
\affiliation{Department of Physics, University of Connecticut, 196 Auditorium Road, Storrs, CT, 06269, USA}
\affiliation{Center for Computational Astrophysics, Flatiron Institute, 162 5th Avenue, New York, NY, 10010, USA}

\author[0000-0003-2630-9228]{Greg L. Bryan}
\affiliation{Department of Astronomy, Columbia University, 550 W 120th Street, New York, NY 10027, USA}
\affiliation{Center for Computational Astrophysics, Flatiron Institute, 162 5th Avenue, New York, NY, 10010, USA}

\author{Romeel Dave}
\affiliation{Institute for Astronomy, University of Edinburgh, Royal Observatory, Edinburgh EH9 3HJ, UK}
\affiliation{Department of Physics \& Astronomy, University of the Western Cape, Cape Town 7535, South Africa}
\affiliation{South African Astronomical Observatories, Observatory, Cape Town 7925, South Africa}

\author{Lars Hernquist}
\affiliation{Center for Astrophysics | Harvard \& Smithsonian, 60 Garden St., Cambridge, MA 02138, USA}

\author{Mark Vogelsberger}
\affiliation{Kavli Institute for Astrophysics and Space Research, Department of Physics, MIT, Cambridge, MA 02139, USA}

\begin{abstract}
The cycle of baryons through the circumgalactic medium (CGM) is important to understand in the context of galaxy formation and evolution. In this study we forecast constraints on the feedback processes heating the CGM with current and future Sunyaev-Zeldovich (SZ) observations. To constrain these processes, we use a suite of cosmological simulations, the Cosmology and Astrophysics with MachinE Learning Simulations (CAMELS), that varies four different feedback parameters of two previously existing hydrodynamical simulations, IllustrisTNG and SIMBA. We capture the dependencies of SZ radial profiles on these feedback parameters with an emulator, calculate their derivatives, and forecast future constraints on these feedback parameters from upcoming experiments.
We find that for a DESI-like (Dark Energy Spectroscopic Instrument) galaxy sample observed by the Simons Observatory all four feedback parameters are able to be constrained (some within the $10\%$ level), indicating that future observations will be able to further restrict the parameter space for these sub-grid models.
Given the modeled galaxy sample and forecasted errors in this work, we find that the inner SZ profiles contribute more to the constraining power than the outer profiles. 
Finally, we find that, despite the wide range of AGN feedback parameter variation in the CAMELS simulation suite, we cannot reproduce the tSZ signal of galaxies selected by the Baryon Oscillation Spectroscopic Survey as measured by the Atacama Cosmology Telescope.
\end{abstract}

\section{Introduction}
Galaxy formation and evolution is a large, unsolved problem of modern extragalactic astronomy. It is a difficult problem to tackle, as multiwavelength observations are needed to fully map the galactic components and the processes occurring in each component. 
An upcoming method to study the questions of galaxy formation and evolution involves using observations and simulations of the circumgalactic medium (CGM). The CGM consists of a large reservoir of multiphase gas surrounding galaxies, extending out to possibly hundreds of kpc. It is a very important component of galactic structure, and is believed to act as a medium through which the flow of material in and out of galaxies cycles (see \citet{Tumlinson2017} for a review). This cycle includes gas falling onto the disks of galaxies from the intergalactic medium (IGM) and CGM, interacting with the central parts of the galaxy to affect processes such as star formation and evolution, and then returning to the CGM and IGM through various feedback channels. An understanding of these feedback mechanisms is extremely important to galaxy formation, but difficult due to the complexity and uncertainty of the baryonic processes affecting the thermodynamics of the gas.

Due to its diffuse nature, the CGM is difficult to observe. Traditionally the main observational technique is through absorption line studies (e.g., \citet{Lanzetta1995,Chen1998,Tumlinson2011,Rudie2012,Tumlinson2013,Werk2014,Chen2018,Lan2018,Zahedy2019,Wilde2021}), but as instruments improve we are able to study it in emission as well (e.g., \citet{Borisova2016,Emonts2016,Ginolfi2017,ArrigoniBattaia2018,Wisotzki2018,Leclercq2020,Zabl2021}). Additionally, an emergent method to study the CGM is with the observation of the cosmic microwave background (CMB). Secondary anisotropies measured in the temperature of the CMB due to the Sunyaev-Zeldovich (SZ) effect \citep{SZ1970} can yield information on the CGM's pressure and density through the thermal (tSZ) and kinetic (kSZ) SZ effects, respectively \citep{SZ1972,SZ1980}. A few examples of recent results include tSZ observations of local galaxies \citep{Pratt2021}, joint tSZ and kSZ cross-correlation measurements of massive galaxies at $z\sim0.5$ \citep{Schaan2021}, and joint X-ray and tSZ observations \citep{Singh2021}.

The tSZ effect describes the increase in CMB photon energies due to inverse Compton scattering off ionized electrons in galaxies and galaxy clusters. This effect produces distortions in the blackbody spectrum of the CMB as a function of frequency with an amplitude proportional to the line-of-sight (LOS) integral of the electron pressure. The tSZ effect equation has the form
\begin{equation}
    \begin{gathered}
    \frac{\Delta T(\nu)}{T_{\text{CMB}}} = f(\nu)y(\theta) \, , \\ 
    y(\theta)= \frac{\sigma_T}{m_{e}c^2}\int_{\text{LOS}}P_{e}(\sqrt{l^2+d_A^2(z){|\theta|}^2})dl \label{eq:tSZ}
    \end{gathered}
\end{equation}
where $\Delta T(\nu)$ is the shift in temperature measured as the tSZ signal, $T_{\rm CMB}$ is the temperature of the CMB, $f(\nu) = x\text{coth}(x/2)-4$ is the spectral function with $x=\frac{h\nu}{k_{B}T_{\text{CMB}}}$,  $h$ is the Planck constant, $k_B$ is the Boltzmann constant, $y(\theta)$ is the Compton-y parameter measured within angular aperture $\theta$, $\sigma_T$ is the Thomson scattering cross section, $m_e$ is the electron mass, c is the speed of light, $d_A(z)$ is the angular diameter distance at redshift $z$, and $P_e$ is the electron pressure.

The kSZ effect describes the Doppler shift of CMB photons scattering off free electrons in galaxies and clusters with nonzero peculiar velocity with respect to the CMB rest-frame. This causes Doppler shifts in the CMB temperature that are directly related to the peculiar momentum, and are proportional to the LOS integral of the peculiar velocity multiplied by the electron number density. The kSZ effect equation has the form 
\begin{equation}
    \begin{gathered}
    \frac{\Delta T}{T_{\text{CMB}}} = \frac{\sigma_T}{c}\int_{\text{LOS}}e^{-\tau(\theta)}n_{e}v_{p}dl \, , \\ \tau(\theta)=\sigma_T \int_{\text{LOS}}n_e(\sqrt{l^2+d_A^2(z){|\theta|}^2})dl \label{eq:kSZ}
    \end{gathered}
\end{equation}
where $n_e$ is the electron number density, $v_p$ is the peculiar velocity, and $\tau(\theta)$ is the optical depth. Certain kSZ estimators allow for the separation of $n_e$ from $v_p$ via the cross-correlation of the CMB with a reconstructed peculiar velocity field from a galaxy survey \citep[e.g.,][]{Ho2009}. Significant kSZ measurements using these cross-correlation estimators were made using spectroscopic galaxy samples \citep{Schaan2016,Schaan2021} and used to measure the average electron density profile of these galaxies \citep{Amodeo2021}. Thus, for this paper we will consider the density profile as the observable from the kSZ signal, which alone already probes the CGM \citep[e.g.,][]{Shao2016}. 

Combining the tSZ and kSZ measurements results in complete thermodynamic information of the CGM \citep{Battaglia2017} that we can use to provide constraints on the physical processes governing galaxy evolution. While we are in an era of increasingly high resolution observations, analytic models and simulations have become an integral part of developing and testing theory against observations. There has been significant progress in studying galaxy formation and evolution (including the CGM) using analytic models \citep[e.g.,][]{Voit2017,Stern2020,Faerman2020,Fielding2021} and hydrodynamical simulations
\citep[e.g.,][]{Oppenheimer2008,Oppenheimer2010,Ford2013,Suresh2017,Fielding2017,Hummels2017,Angles-Alcazar2017b, Oppenheimer2018,Hafen2019,Vogelsberger2020,Fielding2020}, including recent attempts to enhance resolution in CGM simulations \citep[][]{VandeVoort2019,Peeples2019,Hummels2019,Suresh2019}. 

Recent hydrodynamical simulations are able to produce halos with realistic CGM properties compared to observations \citep[e.g.,][]{Nelson2018b,Peeples2019,Nelson2020,Byrohl2021}, but
phenomenological models are required to account for astrophysical processes that occur on scales not resolved by the simulations, such as star formation, supernovae, and supermassive black holes (SMBH).
Thus, there has been much recent work testing these sub-grid physical processes and how they result in properties we can observe, such as the thermodynamics of the CGM \citep{Amodeo2021,Schaan2021,Lim2021,Kim2021}. 

An ambitious effort to this end has begun through the Cosmology and Astrophysics with MachinE Learning Simulations (CAMELS) project \citep{Camels2021,Villaescusa-Navarro2022}. CAMELS is a large suite of simulations varying astrophysical and cosmological parameters, providing an exploratory environment to show the effects of each set of parameters on halo properties. The suite includes both N-body simulations and (magneto)hydrodynamical simulations varying the cosmology and sub-grid physics of the IllustrisTNG \citep{Marinacci2018,Naiman2018,Pillepich2018,Springel2018, Nelson2018a,Nelson2019} and SIMBA \citep{Dave2019} simulations.  Feedback processes within galaxies and halos are poorly understood, so the CAMELS suite provides simulations to explore the parameter space of all of these quantities within two well-established simulation frameworks. 

In this paper, we aim to study how different models and amplitudes of feedback affect the thermodynamics of the CGM using the CAMELS simulations, and whether future observations can be used to further constrain sub-grid feedback models. We discuss our methods in Section~\ref{sec:methods}, which include the calculation of three-dimensional thermodynamic profiles and projection of simulated profiles into SZ profiles. We describe our results in Section~\ref{sec:results}, including an exploration of the constraining power of future profile observations in Section~\ref{sec:results_constraints} and the comparison of the CAMELS simulated profiles to observed profiles in Section~\ref{sec:results_obs}.  A discussion and conclusions are presented in Section~\ref{sec:conclusions}.

\section{Methods}\label{sec:methods}
Here we provide more details on the simulations we use in Section~\ref{sec:methods_simulations}, along with our methods for calculating profiles in Section~\ref{sec:methods_profiles}. We describe the methods for deriving the profile emulators in Sections~\ref{sec:methods_emulator} and~\ref{sec:CMASS_methods}, and how we use the emulator to calculate derivatives (Section~\ref{sec:methods_derivative}) to constrain the feedback processes through a Fisher analysis in Section~\ref{sec:methods_fisher}.

\subsection{Simulations}\label{sec:methods_simulations}
The CAMELS suite contains multiple thousands of simulations, all varying different properties. Each simulation contains 34 snapshots from redshifts $z=6$ down to $z=0$ and has a comoving volume of $(25 {\rm Mpc}/h)^3$. We note that this is a smaller box compared to the volumes of the original IllustrisTNG and SIMBA simulations which have volumes of $(\sim100 {\rm Mpc}/h)^3$, and as such we are limited by not having many high mass objects. 
The impacts of these limitations are discussed below when relevant, and further discussion of the overall limitations of CAMELS can be found in \citet{Camels2021}. Halo and subhalo catalogs are generated using FOF \citep{Huchra1982,Davis1985} and SUBFIND \citep{Springel2001} algorithms, similarly to the original simulations.

The N-body simulations of CAMELS vary the cosmological parameters $\Omega_m$ and $\sigma_8$, and the hydrodynamical simulations additionally vary four astrophysical feedback parameters. We do not consider the impact of $\Omega_m$ and $\sigma_8$ on the CGM, since they will be highly sub-dominant on such scales. The astrophysical parameters represent the amplitudes of the stellar and active galactic nuclei (AGN) feedback models with respect to the corresponding model of the original IllustrisTNG and SIMBA simulations. There are two stellar feedback parameters, $A_{SN1}$ and $A_{SN2}$, along with two AGN feedback parameters, $A_{AGN1}$ and $A_{AGN2}$. More detailed descriptions for each parameter can be found below, but in general they refer to amplitudes of processes such as galactic winds and kinetic-mode black hole feedback implementations. Each of the parameters being varied is simply an amplitude multiplied by the baryonic models to either increase or decrease the amount of feedback, with respect to the amount in the original IllustrisTNG or SIMBA simulation (i.e., the simulations with $A_{SN1}$, $A_{SN2}$, $A_{AGN1}$, $A_{AGN2} = 1.0$ have the same amount of feedback as the original simulation). As noted in \citet{Camels2021}, we emphasize that while the same names and values of the different feedback parameters are used for both CAMELS-IllustrisTNG and CAMELS-SIMBA, the definitions and implementations of the baryonic feedback models differ between the suites corresponding to their respective original simulation versions.

Of all the simulations in the CAMELS suite, we focus on the \textit{1P} set, which is a group of 61 simulations with the same initial random seed and varying only one of the six parameters at a time. For our purposes of exploring how the feedback parameters affect the CGM, we focus on the subset of the \textit{1P} simulations in particular; these vary the four astrophysical parameters, $0.25 \leq (A_{SN1},A_{AGN1}) \leq 4.00$, and $0.5 \leq (A_{SN2},A_{AGN2}) \leq 2.0$ (see Table~\ref{tab:emu_properties}). When the parameters are not being varied, they are held constant at the fiducial amplitude of 1.0. The cosmological parameters in the \textit{1P} simulations are $\sigma_8 = 0.8$, $\Omega_m = 0.30$, and $h = 0.6711$ and are adopted throughout this work.

We emphasize that the original IllustrisTNG and SIMBA simulations implement different models of feedback and use different methods of solving the (magneto)hydrodynamical equations, described further below.

\subsubsection{IllustrisTNG}
IllustrisTNG, following the original Illustris simulations \citep{Vogelsberger2014,Vogelsberger2014b,Genel2014},
uses the AREPO moving-mesh code \citep{Springel2010,Weinberger2020} and includes many physical models such as gas cooling, star formation and evolution, magnetic fields, and feedback from galactic winds, supernovae, and SMBH. The stellar feedback parameters being varied in the CAMELS simulations, $A_{SN1}$ and $A_{SN2}$, correspond to normalization parameters of the total energy injection rate per unit star formation, and the wind speed, respectively. These quantities are functions of redshift, metallicity, and other factors described in more detail in \citet{Pillepich2018}. 

The AGN feedback parameters, $A_{AGN1}$ and $A_{AGN2}$, correspond to aspects of IllustrisTNG's low-accretion kinetic-mode black hole (BH) feedback model. The $A_{AGN1}$ parameter multiplies the feedback energy model, which is a function of gas density, star formation density threshold, and the BH accretion rate. The $A_{AGN2}$ parameter multiplies the minimum injection energy after the BH has accreted enough material to have a feedback event, which is a function of the dark matter velocity dispersion and enclosed mass within the feedback radius. Higher values for $A_{AGN2}$ mean fewer but more energetic feedback events. See \citet{Weinberger2017,Pillepich2018,Camels2021} for the full equations for each of these parameters.

\subsubsection{SIMBA}
SIMBA, following the original MUFASA simulations \citep{Dave2016}, uses the GIZMO meshless finite mass code \citep{Hopkins2015, Hopkins2017} and similarly includes many physical models such as radiative cooling, photoionization heating, star formation and evolution, dust life cycle, and feedback from galactic winds, supernovae, and SMBH. The galactic wind feedback parameter values in SIMBA are based on the Feedback in Realistic Environments (FIRE) simulations \citep{Hopkins2014}. The $A_{SN1}$ parameter modifies the overall amplitude of the mass loading factor, which scales with stellar mass, following \citet{Angles-Alcazar2017b}. Similarly to IllustrisTNG, the $A_{SN2}$ parameter quantifies the amplitude of the wind velocity, although it is computed differently using the circular velocity of the galaxy, following \citet{Muratov2015}. The AGN feedback parameters describe kinetic BH feedback in which gas particles are ejected along the angular momentum axis. The $A_{AGN1}$ parameter adjusts the total momentum flux, which is a function of the bolometric luminosity of the AGN following \citet{Angles-Alcazar2017a}, and the $A_{AGN2}$ parameter affects the maximum jet velocity of the feedback event. See \citet{Dave2019} and \citet{Camels2021} for further details and equations for each of the parameters.

\subsection{Simulated Thermodynamic Profiles}\label{sec:methods_profiles}
Here we describe the process of extracting halo information from the simulations to construct three-dimensional radial profiles and projecting them into observable SZ profiles.

\subsubsection{Three-dimensional Profiles}
\begin{figure*}
    \centering
    \includegraphics[scale=0.65]{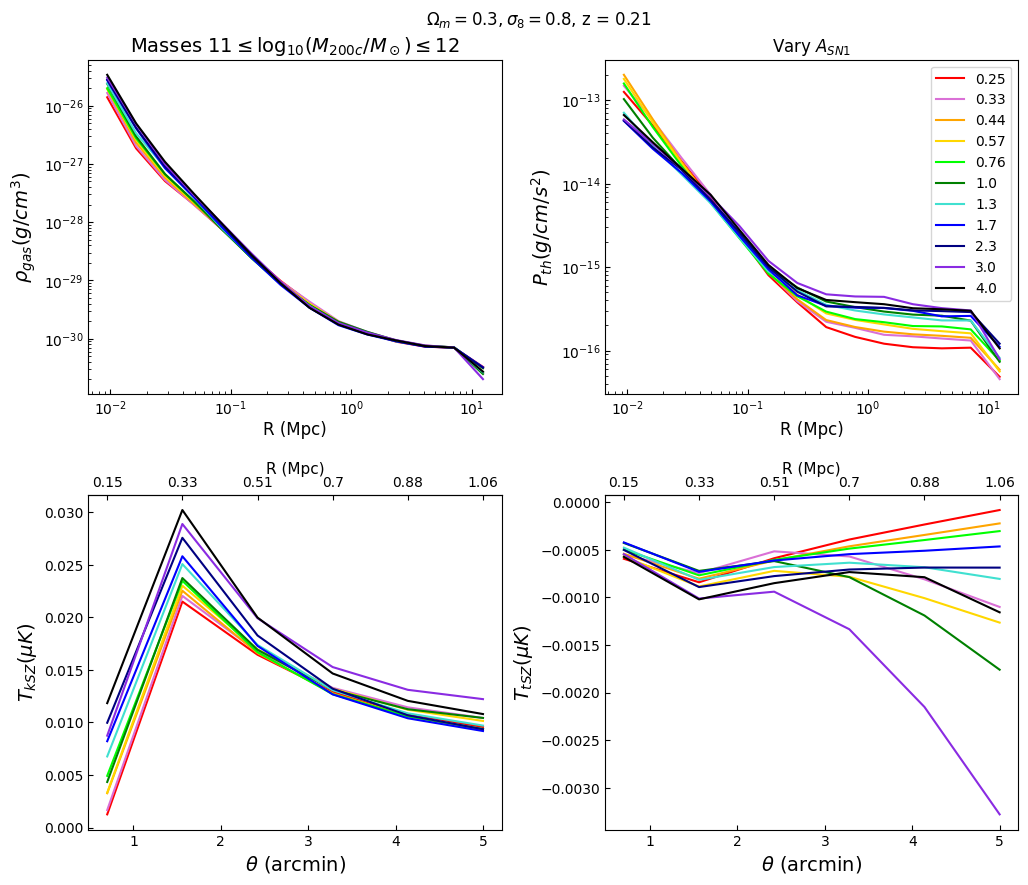}
    \caption{This figure shows 3D profiles produced by \texttt{illstack\char`_CAMELS} (top row) and 2D profiles produced by \texttt{Mop-c-GT} (bottom row) for the subset of CAMELS \textit{1P} simulations varying the $A_{SN1}$ feedback parameter of the SIMBA framework. These are median profiles for the mass range $11 \leq \log_{10}(M_{200c}/M_\odot) \leq 12$ at redshift $z=0.21$. The corresponding  radius values in Mpc are shown on the top axes of the bottom row for straightforward comparison. The values for the $A_{SN1}$ parameter are shown in the legend, with a value of 1.0 being the same amount of feedback as in the original SIMBA simulation. Differences can clearly be seen in all the profiles due to the variation of this particular sub-grid parameter.}
    \label{fig:profiles_3d}
\end{figure*}

We follow the process of \citet{Moser2021} in the extraction of simulated halo information and construction of three-dimensional thermodynamic profiles. We use the repository, \texttt{illstack\char`_CAMELS}\footnote{\url{https://github.com/emilymmoser/illstack_CAMELS}} (a CAMELS-specific version of the original, more general code \texttt{illstack} used in \citet{Moser2021}) to extract and stack the halo gas density and pressure information from the simulations and to create three-dimensional radial density and pressure profiles for the radial range of $\sim 0.01-10$ Mpc. We note that this radial range was chosen to cover the extent of SZ observations, but much of the two-halo contribution at higher radii will be similar for each profile, due to the smaller simulation volumes. Therefore, if only one-halo profiles were desired one could simply only use the values at lower radii. An example of simulated median density and pressure profiles can be seen in the top row of Figure~\ref{fig:profiles_3d} for the halos of the CAMELS-SIMBA suite with masses $11 \leq \log_{10}(M_{200c}/M_\odot) \leq 12$ at redshift $z=0.21$ varying the $A_{SN1}$ feedback parameter. For this particular feedback parameter, mass range, and redshift, the outer region of the density profile does not seem to be affected by the variation of the feedback as much as the inner profile, where increasing the mass loading factor can be seen to increase the density. However, the variation of feedback amplitude can be seen to have a large effect in both radial limits of the pressure profile. Increasing the mass loading factor results in an overall increase in amplitude of the outer pressure profile, while the spread in inner profile is less monotonic. The bottom row of the figure shows these profiles projected into two-dimensional SZ profiles, described further below.

\subsubsection{Two-dimensional Profiles}
The signals of the kSZ and tSZ effects are integrated along the LOS, so we project our 3D profiles into 2D profiles of observable quantities similarly. We follow the process described in \citet{Moser2021}, briefly summarized below. We use the repository \texttt{Mop-c-GT}\footnote{\url{https://github.com/samodeo/Mop-c-GT}} (Model-to-observable projection code for Galaxy Thermodynamics), introduced in \citet{Amodeo2021}, to project the simulated thermodynamic profiles into profiles of temperature shifts in the CMB signal due to the kSZ and tSZ effects, shown in Equations~\ref{eq:tSZ} and ~\ref{eq:kSZ}. The 2D profiles are computed from the 3D profiles by projecting 10 Mpc along the line-of-sight in front of and behind the halos. Then Mop-c-GT forward-models these 2D profiles to experiment-specific SZ signals through beam convolution and aperture photometry. See \citet{Amodeo2021,Schaan2021,Moser2021} for further details on the kSZ reconstruction and other systematic uncertainties of \texttt{Mop-c-GT}. 

An example of projected SZ profiles produced by \texttt{Mop-c-GT} can be seen in the bottom row of Figure~\ref{fig:profiles_3d}, with density profiles projected into kSZ profiles shown on the left, and pressure profiles projected into tSZ profiles at a frequency of 150 GHz shown on the right. We note that at this frequency the tSZ effect results in a temperature decrement, thus we have negative temperature values. These profiles have been projected with a Gaussian beam (see Section~\ref{sec:CMASS_methods} for more details on beams) simply to show an example of the simulated SZ profiles we are able to calculate and study.

\subsection{CAMELS Profile Emulator}\label{sec:methods_emulator}
Previously in \citet{Moser2021}, the profile differences of various samples and models were quantified by fitting to a generalized Navarro-Frenk-White \citep[GNFW;][]{gnfw} profile. This process required prior knowledge of the GNFW parameters obtained from previous simulations \citep{Nagai2007, Battaglia2012a,Battaglia2016}, and further prior knowledge along with manual experimentation to know which parameters to leave free versus fixed in the Markov chain Monte Carlo (MCMC) fits. For the study presented in \citet{Moser2021} this kind of parametric model fitting was achievable and effective; however, the current study aims to look at different profiles varying many more parameters relevant to sub-grid physical processes, masses, and redshifts, so the rigidity of parametric model fitting was more limiting, especially in low-mass halos where the baryonic effects are expected to be significant. Therefore, we developed a different method to capture the differences of the principal components of the profiles by deriving a \textit{profile emulator}. For the purposes of this study, an emulator is a multidimensional interpolator and acts as a predictive model for the thermodynamic profiles. 

The CAMELS suite provides many simulations varying each of the parameters, but it is still a discrete grid of data points as opposed to a continuous distribution. The emulator can essentially interpolate between the data points (``data points" meaning the profiles of individual simulations, representing a single realization of a combination of parameters) and return profiles for areas in the parameter space not explicitly covered by the simulations. 
To construct the emulator, we follow the process described in \citet{Heitmann2009} and \citet{Cromer2021b}, in which the basis vectors of the profiles are decomposed using principal component analysis (PCA; \citet{Pearson1901}, see \citet{Jolliffe2016} for a review). Following usual PCA, the profile is deconstructed into the principal components that are assigned weights to best match the profile, and the weights are interpolated using a radial basis function (RBF) interpolator. The principal components are sorted by decreasing variance, and the number of components kept is a modeling choice, which will impact the amplitude of residual variance in the analysis. In this study we proceed with 12 components but note that adding more components does not affect the results of the emulator. We use the repository \texttt{emu\char`_CAMELS}\footnote{\url{https://github.com/emilymmoser/emu_CAMELS}} (a CAMELS-specific application of the repository \texttt{ostrich}, used in \citet{Cromer2021b}) to construct an emulator for each 3D profile type (density and pressure) and each feedback parameter ($A_{SN1}$, $A_{SN2}$, $A_{AGN1}$, $A_{AGN2}$) since the \textit{1P} simulations only vary one of the paramters at a time. The median and mean CAMELS profiles are generally different due to a small number of outlier halos, so we can build an emulator for each type of profile as well. The emulator interpolates the profiles it is given, so if a median profile emulator is desired we give it only the median profiles, and similarly for the mean profiles.

We have constructed a general CAMELS profile emulator that is a function of three quantities: feedback parameter, redshift, and mass, described in Table~\ref{tab:emu_properties}. The current capability of the emulator includes redshifts of the ten snapshots available in the range $0 \leq z \leq 0.54$.
We chose four mass bins spanning the range $11 \lesssim \log_{10}(M/M_{\odot}) \lesssim 13$ which were chosen to have a large enough sample of halos to have stable trends in the profiles. As previously mentioned, the small box sizes of the CAMELS simulations limit our mass selections to objects with halo masses  $M_{200c}\lesssim 10^{13} M_{\odot}$, and the higher mass bins have significantly fewer objects than the lower mass bins.

\begin{figure}
    \centering
    \includegraphics[scale=0.6]{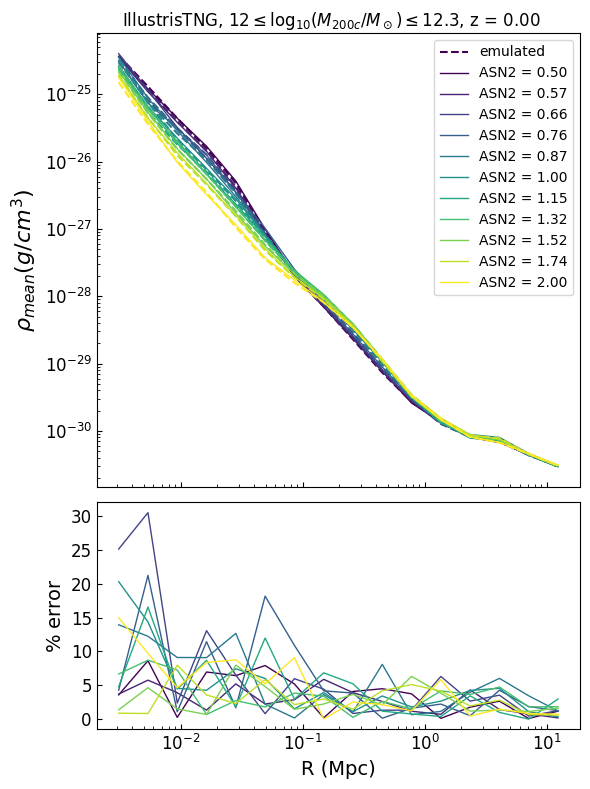}
    \caption{This figure shows an example of the emulator's accuracy for mean density profiles of the CAMELS-IllustrisTNG suite varying feedback parameter $A_{SN2}$ at redshift $z=0.0$. In the top panel, the profiles taken from the simulations are shown by solid curves, with the different colors representing a different $A_{SN2}$ value, described by the legend. The emulator's prediction for each value is shown by the dashed curves of the same color. The bottom panel shows the percent error for each emulated feedback value as a function of radius, showing that for this particular sample, we are able to achieve $\lesssim10\%$ accuracies for most of the radial range.
    }
    \label{fig:emu_accuracy}
\end{figure}

A demonstration of the general CAMELS emulator's accuracy in shown in Figure~\ref{fig:emu_accuracy} with mean density profiles from the CAMELS-IllustrisTNG suite. These profiles were calculated for halos at redshift $z=0.0$ within the third mass bin $12.0 \leq \log_{10}(M_{200c}/M_\odot) \leq 12.3$ (see Table~\ref{tab:emu_properties}) varying the $A_{SN2}$ parameter. The top panel in this figure shows the simulated profiles produced by \texttt{illstack\char`_CAMELS} in solid lines, with the different colors corresponding to a different $A_{SN2}$ feedback value, described by the legend. The dashed lines show the emulator's predictions for each of the feedback values as if it did not have the information for the given profile (i.e., the profile has been dropped when constructing the emulator), and the bottom panel shows the percent errors as a function of radius. For this particular example, the emulator is able to achieve $\lesssim10\%$ accuracy of the profiles for most of the radial range (and even lower for the outer radial points).

\begin{table*}
    \centering
    \begin{tabular}{c|c|c|c|c|c}
        \multicolumn{2}{c|}{Feedback Parameters} & \multicolumn{2}{c|}{Mass Bins} & \multicolumn{2}{c}{Redshifts} \\
        \hline
        $A_{SN1}$ $\&$ $A_{AGN1}$ & $A_{SN2}$ $\&$ $A_{AGN2}$ & CAMELS & CMASS & CAMELS & CMASS \\
        \hline
        0.25, 0.33, 0.44, 0.57, & 0.50, 0.57, 0.66, 0.76, & 11.0-11.5-12.0- & 12.12-13.98* & 0.00, 0.05, 0.10, 0.15, & 0.54 \\
        0.76, 1.00, 1.32, 1.74, & 0.87, 1.00, 1.15, 1.32, & 12.3-13.1 & & 0.21, 0.27, 0.33, & \\
        2.30, 3.03, 4.00 & 1.52, 1.74, 2.00 & & & 0.40, 0.47, 0.54 & \\
        \hline
    \end{tabular}
    \caption{Values of the three properties interpolated by the emulators. The left two columns show the values for each feedback parameter, the middle two columns show the mass bins for the general CAMELS profile emulator and the CMASS-specific emulator in $\log_{10}(M_{200c}/M_\odot)$ (*we note that halos with masses $\gtrsim 10^{13.1} M_\odot$ are missing from the emulators due to the smaller volumes of the CAMELS simulations), and the right columns show the redshifts included for both emulators. See Sections~\ref{sec:methods_emulator} and~\ref{sec:CMASS_methods} for further detail.}\label{tab:emu_properties} 
\end{table*}

\subsection{CMASS Emulator}\label{sec:CMASS_methods}
We construct an ``observed" galaxy sample from which we derive an emulator to model a CMASS-like (constant stellar mass) sample with a DESI-like \citep[Dark Energy Spectroscopic Instrument;][]{DESI2016} number of galaxies. The CMASS sample was observed by the Baryon Oscillation Spectroscopic Survey \citep[BOSS, Data Release 10;][]{Ahn2014}, and contains hundreds of thousands of luminous red galaxies (LRGs) with a median redshift of $z=0.55$. SZ profile measurements have already been made for the CMASS sample \citep{Amodeo2021,Schaan2021}, which was chosen to contain galaxies in the region covered by the Atacama Cosmology Telescope (ACT). DESI will improve upon these statistics by increasing the number of LRGs observed to be on the order of several millions \citep{DESI2016,Levi2019}. Therefore,  we model a sample of halos to match the physical characteristics of CMASS (further described below) while forecasting signals for a larger survey of galaxies, such as DESI.

A result of \citet{Moser2021} was that certain modeling choices of the sample are important to consider when interpreting the observed profiles, particularly the mass-distribution-matching of the observed sample. Therefore, to appropriately model the CMASS sample, we build an emulator of halo mass-weighted profiles following the same process described in detail in \citet{Moser2021}. As the CMASS sample is peaked around the median redshift of $z=0.55$, we use profiles only from the corresponding simulation snapshots. Since we are using profiles from a single redshift and no longer interpolating over mass (due to the mass distribution weighting) the emulator is reduced to a one-dimensional interpolation over feedback parameters. We refer to this emulator as the CMASS emulator, but specify that it is actually an emulator for a CMASS-like sample in properties such as the mass distribution, but DESI-like in terms of the better statistics of a larger number of observed objects.

Since \texttt{Mop-c-GT} forward-models the profiles as observed by individual experiments, we can use it to test the constraining power of different beams, e.g. an ideal case of a Gaussian beam compared to a more realistic beam with tails. When emulating CMASS profiles from ACT we convolve the profiles with the ACT beam, which has non-Gaussian and scale-dependent profiles \citep{Naess2020,Amodeo2021,Schaan2021}. We have the observed SZ profiles and error bars from \citet{Amodeo2021,Schaan2021} (which have been convolved with the ACT beam) and we can compare these profiles (and the constraining power of the profiles) with forecasted profiles of a future experiment, such as for the CMASS/DESI-like sample observed by the Simons Observatory (SO) \citep{SO2019}. In this case, we would convolve the profiles with a Gaussian beam to show the best quality of observations possible for this experimental setup. We model the Gaussian beam to have an angular resolution of $1.4^\prime$ at frequency 150 GHz to match the forecasted SO experimental setup described in \citet{SO2019}. Since we do not have actual error bars for the SO experiment, we use the forecasted errors as derived in \citet{Battaglia2017}, in which the authors use a semi-analytical foreground model including contributions from primary CMB fluctuations, extragalactic radio emission, and the cosmic infrared background.

\subsection{Derivatives of Feedback Models}\label{sec:methods_derivative}
We quantify how the profiles change as a function of feedback parameter by computing numerical derivatives. We choose a fiducial value for each feedback parameter, $A_0$, chosen to be 1.0, as it is the middle value for each feedback parameter being varied (see Table~\ref{tab:emu_properties}) and equivalent to the fiducial amplitudes in the original IllustrisTNG and SIMBA simulations. Then we use the emulator to calculate profiles with $A = A_0\pm \Delta A$, where $\Delta A$ is some small value in the feedback parameter range. We chose $\Delta A$ to be 0.1 for $A_{SN1}$ and $A_{AGN1}$ and 0.05 for $A_{SN2}$ and $A_{AGN2}$ to have step sizes small enough to be within the simulation range but large enough to show differences between profiles. Then we calculate numerical derivatives of the emulated profiles of the form
\begin{equation}
    \frac{df}{dA} \approx \frac{f(A^+)-f(A^-)}{2\Delta A} \label{eq:derivative}
\end{equation}
where $f$ is the profile calculated by the emulator given a feedback parameter $A$, $\Delta A$ is the change in the feedback parameter value being varied around the fiducial point, $A^+$ is equivalent to $A_0 + \Delta A$, and $A^-$ is equivalent to $A_0 - \Delta A$. The associated error term from this approximation is of the order $\mathcal O ({\Delta A}^2$), but is much smaller in magnitude than the errors seen in Figure~\ref{fig:emu_accuracy}.

\subsection{Fisher Analysis}\label{sec:methods_fisher}
Fisher matrix analyses (\citet{Fisher1935}, see \citet{Tegmark1997} for a review) are extremely valuable and widely used in astronomy. They allow for forecasting constraints of future experiments, along with estimating the magnitude of errors necessary for a desired significance of measurement. In general, the process is to combine derivatives of the thermodynamic profiles as a function of feedback parameter with an observed (or forecasted) covariance matrix of the experiment. To forecast the constraints of the CMASS emulator, we use the covariance matrix of forecasted errors for the SO experiment of \citet{Battaglia2017}.

The form of the Fisher matrix is shown in Equation~\ref{eq:fisher}, with the first line showing the kSZ matrix, the second line showing the tSZ matrix, and the last line showing the total Fisher matrix as the combination of the kSZ and tSZ matrices:  
\begin{equation}
    \begin{gathered}
    F_{j,k}^{kSZ} = \sum_{\theta_d,\theta_{d^\prime}} \frac{\partial f(\theta_d)}{\partial p_j} (C_{kSZ}^{-1})_{\theta_d,\theta_{d^\prime}}\frac{\partial f({\theta_{d^\prime}})}{\partial p_k} \\
    F_{j,k}^{tSZ} = \sum_{\theta_d,\theta_{d^\prime}} \frac{\partial f(\theta_d)}{\partial p_j} (C_{tSZ}^{-1})_{\theta_d,\theta_{d^\prime}}\frac{\partial f({\theta_{d^\prime}})}{\partial p_k} \\
    F^{total} = F^{kSZ}+F^{tSZ}
    \end{gathered}\label{eq:fisher}
\end{equation}
where $f(\theta_d)$ is the projected density or pressure profile given an aperture $\theta_d$, $(C^{-1})_{\theta_d,\theta_{d^\prime}}$ is the inverse covariance matrix for the aperture photometry filter being used (we have different covariance matrices for kSZ and tSZ observations, seen by the top and middle lines), and $p_j$ is the $j^{th}$ parameter being forecasted. We note that the total Fisher analysis does not include the covariance between the kSZ and tSZ observations. The joint SZ measurements by \citet{Schaan2021} showed that this covariance is not negligible. However, it is expected that future CMB observations will have sufficiently more frequency coverage that component separation will largely suppress such covariances \citep{SO2019}. We then plot the contours of the constraints in a corner plot using the repository GetDist \citep{Lewis2019}. 

\section{Results}\label{sec:results}
We show the forecasted feedback model constraints of a CMASS/DESI-like sample observed by SO in Section~\ref{sec:results_constraints}, including further exploration of the contribution of different radial components of the profiles. We show the comparison of SZ profiles calculated from the CAMELS simulations to observations in Section~\ref{sec:results_obs}, and derive the combination of parameters resulting in the best fit CAMELS profile to the ACT observations.

\begin{figure*}
    \centering
    \includegraphics[scale=0.7]{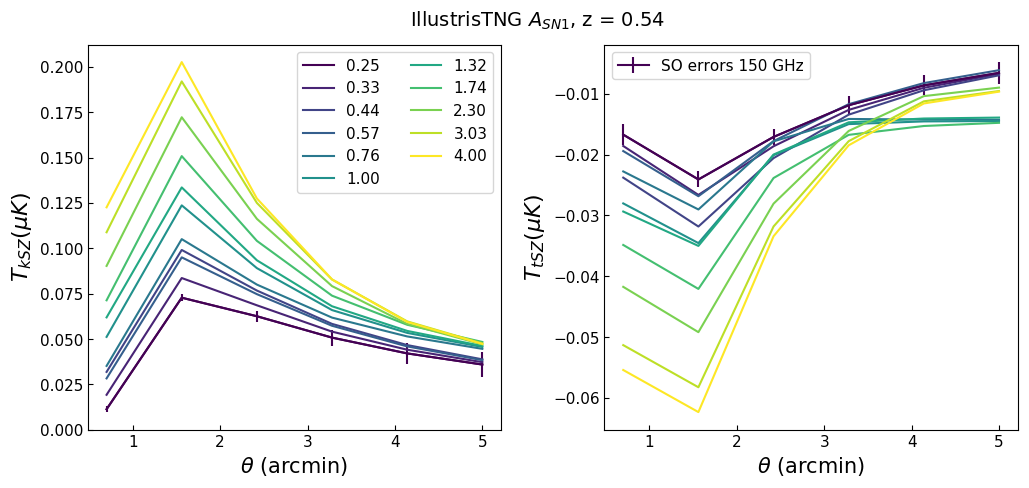}
    \includegraphics[scale=0.7]{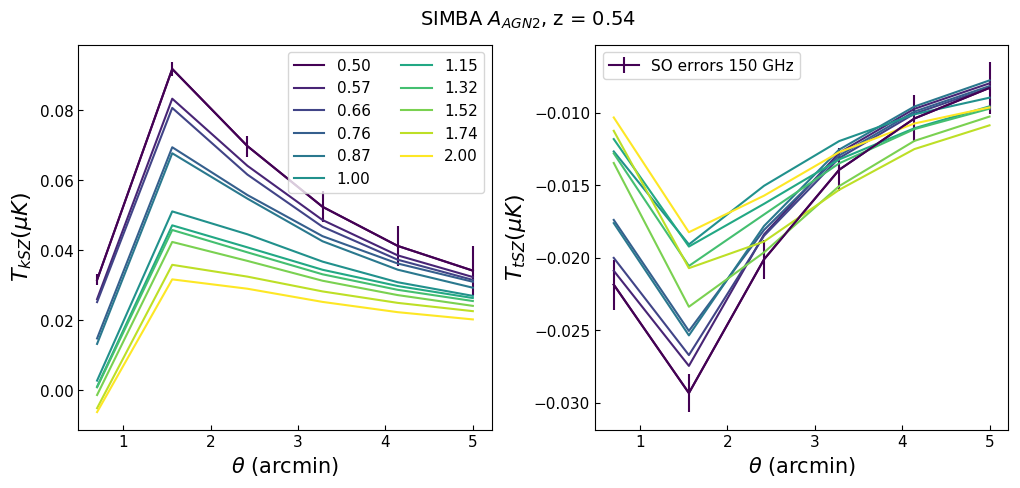}
    \caption{This figure shows kSZ (left) and tSZ (right) projections varying feedback parameters for the CAMELS-IllustrisTNG suite (top, varying $A_{SN1}$) and the CAMELS-SIMBA suite (bottom, varying $A_{AGN2}$). The profiles have been weighted to match the mass distribution of the CMASS sample, limits shown in Table~\ref{tab:emu_properties} and process described in Section~\ref{sec:CMASS_methods}. The forecasted SO error bars are shown in purple in each panel (see Section~\ref{sec:CMASS_methods}), indicating that with this level of sensitivity we will be able to differentiate and constrain these models.
    }
    \label{fig:obs_SO}
\end{figure*}

\subsection{Constraints on Feedback Models with Near-future SZ Observations}\label{sec:results_constraints}
First, we forecast the constraints on the underlying feedback models given observed profiles. In Figure~\ref{fig:obs_SO} we show kSZ and tSZ profiles from the CAMELS-IllustrisTNG suite varying the $A_{SN1}$ feedback parameter in the top row, and SZ profiles from the CAMELS-SIMBA suite varying the $A_{AGN2}$ feedback parameter in the bottom row, as illustrations. These profiles have all been weighted to match the mass distribution of the CMASS sample and convolved with a Gaussian beam of $1.4^\prime$. In all panels, the forecasted SO error bars at 150 GHz are plotted in purple. It can be seen that for most radial bins, the error bars are smaller than the spread of profiles varying the feedback parameters, indicating that with this level of precision, we will be able to constrain these thermodynamic processes. 

\begin{figure*}
    \centering
    \includegraphics[scale=0.7]{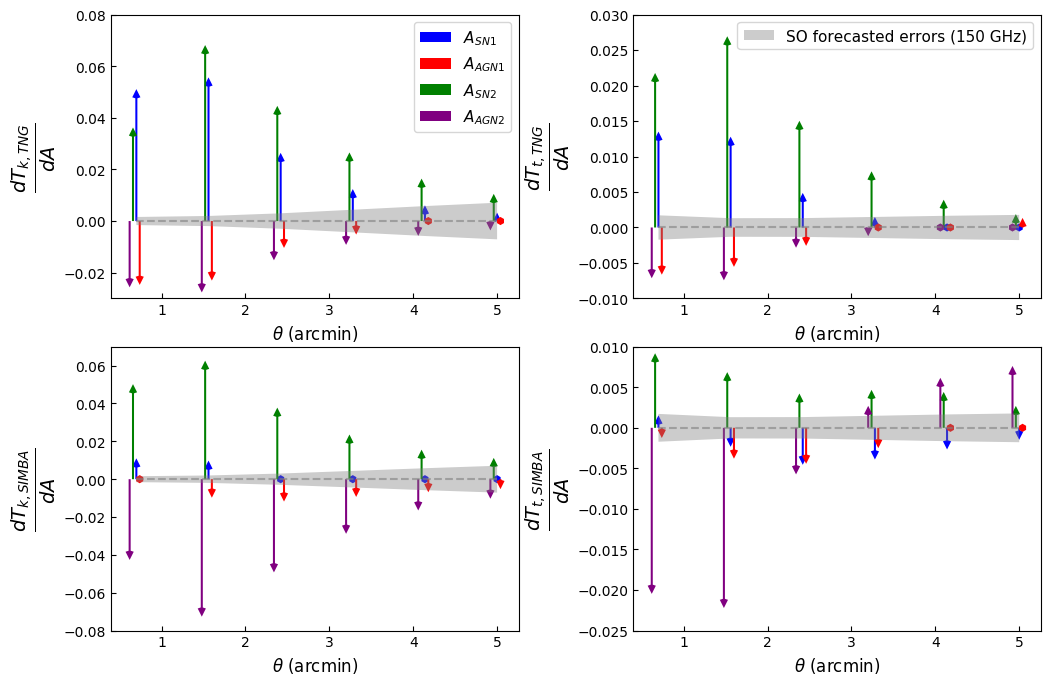}
    \caption{This figure shows the derivatives for each parameter at $A=1.0$ by the colored arrows for kSZ (left) and tSZ (right) profiles of CAMELS-IllustrisTNG (top row) and CAMELS-SIMBA (bottom row). Positive arrows indicate an increase in that particular feedback parameter leads to an increase in amplitude of the profile at that particular radial point. Small derivatives are indicated by a hexagonal point close to zero, and the gray error band in each panel shows the forecasted SO errors of the profiles. This figure shows the emulator for each suite and profile type is sensitive to changes in each of the parameters differently, leading to the different levels of constraint we are able to achieve.     
    }
    \label{fig:parameter_derivs}
\end{figure*}

\subsubsection{Derivatives and Constraints}
To quantify the forecasted constraints, we follow the processes outlined in Sections~\ref{sec:methods_derivative} and~\ref{sec:methods_fisher} using the CMASS emulator. In Figure~\ref{fig:parameter_derivs} we show a visualization of the derivatives for each parameter to display how each is affecting the shape of the radial profiles. The top row of the figure shows the derivatives for the kSZ (left) and tSZ (right) profiles from CAMELS-IllustrisTNG, and the bottom row similarly shows the derivatives for CAMELS-SIMBA profiles. We note that all derivatives were taken at the fiducial value for each feedback parameter, $A=1.0$, and that these values could change if calculated at other points in the parameter space. The different colored arrows show the derivatives by direction and magnitude for each parameter, described by the legend, and the gray band shows the forecasted SO errors at 150 GHz, also shown in Figure~\ref{fig:obs_SO}. We further note that in this figure a positive arrow (arrow pointing up) indicates an increase in the amplitude of the profile for an increase in the given parameter, and a negative arrow (arrow pointing down) means a decrease in the amplitude of the profile for an increase in the given parameter. 
This figure shows that, for each profile and simulation suite, we see a unique combination of effects from the changes in each parameter, in both direction and magnitude, which helps to break degeneracies in the parameter space. In nearly all panels, we see larger derivatives of the $A_{SN2}$ parameter, shown by green arrows, which should lead to tighter constraints (along with $A_{AGN2}$ shown by purple arrows for SIMBA). In nearly all panels we see relatively small derivatives from $A_{AGN1}$, which should lead to wider constraints. 

How well a parameter is predicted to be constrained depends on the ratio of its derivative with respect to projected measurement errors (See Equation~\ref{eq:fisher}). These projected error bands are shown with gray bands in Figure~\ref{fig:parameter_derivs}. It can be seen that the emulator is typically more sensitive to changes in the inner radial region of the SZ observations than the outer. It is important to note that the inner radial ranges of these forecasted measurements probe the outer, largely unconstrained regions of the CGM. These forecasted constraints on the outer region are still more sensitive to changes in the outer profile than X-ray (i.e. X-ray sensitivity falls off much more rapidly with radius). For the kSZ profiles, the errors increase as radius increases (due to primary CMB fluctuations and covariances, see \citet{Schaan2021}), and the derivatives decrease in magnitude as we increase in radius. The errors on the tSZ profiles are significantly smaller than the kSZ errors, but we can also see that the emulator is more sensitive to changes in the profiles for certain radial points.

While the magnitudes of the derivatives do have an effect on the Fisher analysis, they are not the only important details in determining the level of constraint we are able to achieve. Any feature the emulator could identify to differentiate the shapes of the profiles, such as the inflection in the derivatives of the $A_{AGN2}$ parameter in the SIMBA tSZ profiles, could lead to tighter constraints, and any degeneracy in the profiles due to variations of the parameters could lead to wider constraints.

For CAMELS-IllustrisTNG, as we increase $A_{SN1}$ (energy per SFR) the density and pressure also increase, which suggests that $A_{SN1}$ is heating the gas and suppressing SFR, leaving more gas that has not converted to stars. As we increase $A_{AGN1}$ (energy per BH accretion rate) the density and pressure decrease, which suggests that this parameter is pushing gas out into the IGM. For $A_{SN2}$ (wind speed) we see similar but stronger trends than $A_{SN1}$, in which the gas is being built up. For $A_{AGN2}$ (ejection speed/burstiness), we see similar but slightly stronger trends than $A_{AGN1}$.
For CAMELS-SIMBA, as we increase $A_{SN1}$ (mass loading factor), the density is slightly increased in the inner region, and pressure is slightly decreased for most of the radial range. Similarly to IllustrisTNG, the increase in density could be due to less of the gas being converted to stars and thus more gas circulating through the CGM, but for SIMBA this effect does not propagate to larger scales like it does for IllustrisTNG. The decrease in pressure could be due to enhanced cooling, perhaps from the higher gas density and/or higher metallicity since the winds are transporting metals out of the interstellar medium more efficiently. For SIMBA, the primary effect from AGN feedback is to change the amount of hot CGM baryons. As we increase $A_{AGN1}$ (momentum flux) we see similar trends to the changes in $A_{AGN1}$ for IllustrisTNG, suggesting this parameter is also pushing gas out into the IGM. As we increase $A_{SN2}$ (wind speed) we see similar trends to the changes in $A_{SN2}$ for IllustrisTNG, in which the gas is being built up and not converting to stars. It has been shown that for halo masses $\lesssim10^{13} M_\odot$ the AGN feedback in SIMBA tends to strongly remove baryons from halos \citep{Sorini2021}. As we increase $A_{AGN2}$ (jet speed) we see a large decrease in density and large decrease in pressure in inner regions, in qualitative agreement with IllustrisTNG, but the pressure increases in the outer regions. This removal of baryons could be accompanied by another mechanism such as shocks heating the outer profile, resulting in the inflection and increase of pressure. Additionally, jets in SIMBA are collimated \citep{Christiansen2020} which may imply a lower ability to heat up gas directly in the inner region (hence lower pressure) with higher velocity jets, while more efficiently heating gas farther out as the jet cocoon expands. It is possible there are also environmental effects given the long range effect of jet feedback in SIMBA \citep{Borrow2020}, with stronger jets from other halos reaching the outer profile of a given halo.

\begin{figure*}
    \centering
    \includegraphics[scale=0.5]{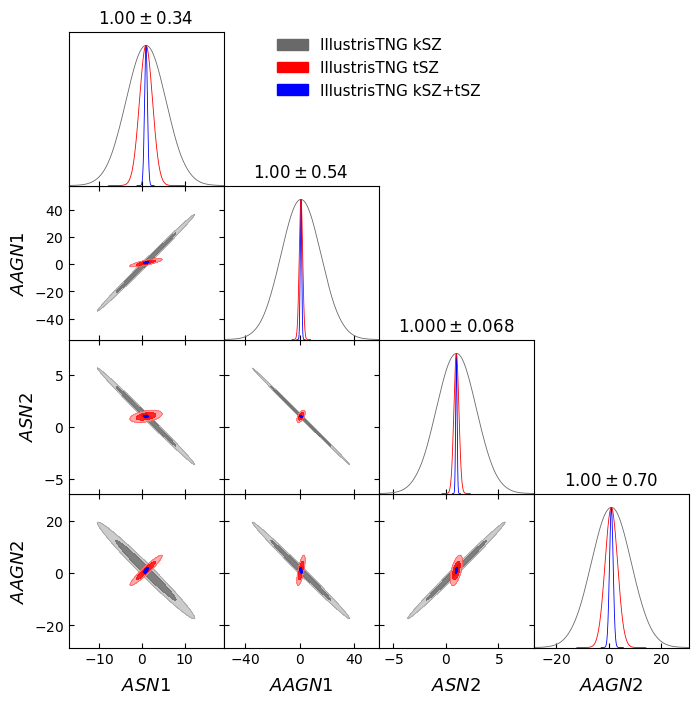}
    \includegraphics[scale=0.5]{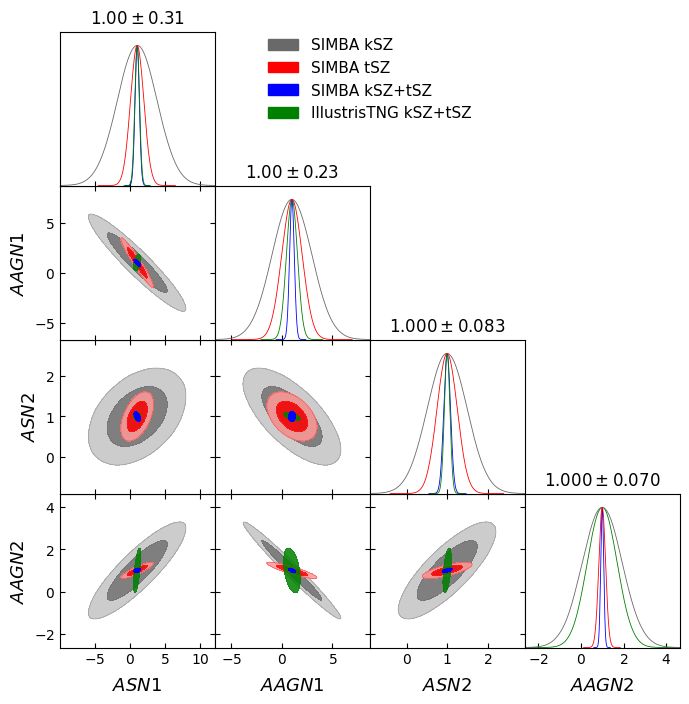}
    \caption{Corner plots showing constraints from the CMASS emulator for CAMELS-IllustrisTNG (left) and CAMELS-SIMBA (right). The different colored contours show the constraint for the feedback parameter using only kSZ (grey), only tSZ (red), and the combination of kSZ and tSZ (blue). Since the degeneracies are strong in the left corner plot for CAMELS-IllustrisTNG, we inlcuded the combined constraints of CAMELS-IllustrisTNG to the CAMELS-SIMBA corner plot (right panel) to show their combined constraints on the same scale. These profiles and derivatives were calculated using a Gaussian beam, and the $\pm 1\sigma$ value of the distribution for each parameter using the combined constraints (blue contours) is shown above each panel on the diagonal and listed in Table~\ref{tab:emu_constraints}. We expect that with future observations we should be able to place constraints within the $10\%$ level on these sub-grid models.}
    \label{fig:corner_emu_CMASS}
\end{figure*}

Figure~\ref{fig:parameter_derivs} can be understood in tandem with Figure~\ref{fig:corner_emu_CMASS}, in which we combine the derivatives into a Fisher matrix to compute the forecasted constraints. The left corner plot shows the forecasted constraints for CAMELS-IllustrisTNG and the right plot shows the constraints for CAMELS-SIMBA. Like the derivatives, we see different levels of constraints for each parameter and differences between the two CAMELS suites in general. First, in the left corner plot for CAMELS-IllustrisTNG we see strong degeneracies among the parameters from individual measurements, particularly for kSZ (gray contours). This is not entirely unexpected, as the parameters being varied in the CAMELS simulations are only amplitudes of the feedback processes (see Section~\ref{sec:methods_simulations} and references therein for definitions), and as such could result in the same profile shape only varying in amplitude. Additionally, it can be seen that the tSZ measurements provide tighter constraints than the kSZ measurements, which is expected due to the smaller tSZ error bars seen in Figures~\ref{fig:obs_SO} and~\ref{fig:parameter_derivs}. In order to break the degeneracies to result in usable constraints, we need to combine both kSZ and tSZ measurements which can be seen by the blue contours. We are able to achieve a $34\%$ constraint for $A_{SN1}$, $54\%$ constraint for $A_{AGN1}$, $6.8\%$ constraint for $A_{SN2}$, and $70\%$ constraint for $A_{AGN2}$, all using a Gaussian beam. The more realistic ACT beam yields slightly wider constraints, see Table~\ref{tab:emu_constraints} for the direct comparison. As shown by the derivatives in Figure~\ref{fig:parameter_derivs}, these profiles are most sensitive to changes in the $A_{SN2}$ parameter, resulting in the tightest constraint, and least sensitive to changes in the $A_{AGN2}$ parameter.

The right corner plot of Figure~\ref{fig:corner_emu_CMASS} shows the constraints for the profiles of the CAMELS-SIMBA suite. We do not see the same level of degeneracy among the parameters as for the IllustrisTNG profiles, which could lead to the overall better constraints. We are able to achieve a $31\%$ constraint for $A_{SN1}$, $23\%$ constraint for $A_{AGN1}$, $8.3\%$ constraint for $A_{SN2}$, and $7.0\%$ constraint for $A_{AGN2}$. These profiles are most sensitive to changes in the $A_{SN2}$ and $A_{AGN2}$ parameters (seen by the derivatives of Figure~\ref{fig:parameter_derivs}), but the others are also well-constrained.

We note again that the physical meanings of the four feedback parameters are different between the CAMELS-IllustrisTNG and CAMELS-SIMBA suites, so we cannot directly compare the constraints of the corresponding parameters. However, we can make the comparison of the overall constraining power of the emulators for each suite. The CMASS emulator of the SIMBA profiles is able to achieve tighter constraints on each of the parameters (aside from $A_{SN2}$) compared to the CMASS emulator of the IllustrisTNG profiles, indicating the SIMBA profiles are more sensitive to the changes in these parameters than the IllustrisTNG profiles. It is well-known within CAMELS that SIMBA has a stronger AGN feedback implementation, while for IllustrisTNG the AGN feedback has a milder effect in general.

All profiles used in Figures~\ref{fig:parameter_derivs} and~\ref{fig:corner_emu_CMASS} were computed with a Gaussian beam, but we also forecasted the constraints using the ACT beam (see Section~\ref{sec:CMASS_methods}) to see exactly how much constraining power we lose due to the imperfections of the beam. We list the constraints in Table~\ref{tab:emu_constraints} derived by using both beams for direct comparison; it can be seen that the Gaussian beam yields tighter constraints for each suite and parameter, which is expected due to the higher signal-to-noise of the profiles and level of precision.

\begin{table*}
    \centering
    \begin{tabular}{c|c|c|c|c|c|c|c|c}
         \multicolumn{9}{c}{CMASS Emulator $\%$ Constraints}  \\
         \hline
          Simulation & \multicolumn{4}{c|}{IllustrisTNG} & \multicolumn{4}{c}{SIMBA} \\
          \hline
          Beam Model & \multicolumn{3}{c|}{Gaussian} & ACT & \multicolumn{3}{c|}{Gaussian} & ACT \\ 
          \hline
          Radial Range& Inner & Outer & Total & Total & Inner & Outer & Total & Total \\
         \hline
        $A_{SN1}$ & 41 & 89 & 34 & 47 & 35 & UC & 31 & 41 \\
        $A_{AGN1}$ & 95 & UC & 54 & 76 & 27 & UC & 23 & 46 \\
        $A_{SN2}$ & 10 & 35 & 6.8 & 10 & 10 & 21 & 8.3 & 10 \\
        $A_{AGN2}$ & UC & UC & 70 & 81 & 8.2 & 85 & 7.0 & 8.9 \\
        \hline
    \end{tabular}
    \caption{Constraints on each parameter from the CMASS emulator comparing the ideal Gaussian beam and the ACT beam (see Section~\ref{sec:CMASS_methods}) and the constraints of the different radial components (see Section~\ref{sec:results_constraints_split}). ``Total'' means the entire profile is used in the analysis, and the $\%$ constraints are the $1.0\pm 1\sigma$ confidence of the blue contours (combined kSZ and tSZ) shown in Figure~\ref{fig:corner_emu_CMASS} for the Gaussian beam. The ``inner'' and ``outer'' constraints of the Fisher analysis using the Gaussian beam are shown in Figure~\ref{fig:constraint_split}. Unconstrained parameters with percentages larger than 100 have been indicated as ``UC''. }\label{tab:emu_constraints} 
\end{table*}

\subsubsection{Radial Information Content}\label{sec:results_constraints_split}
We further explore the forecasted feedback parameter constraints by calculating which radial component of the profile is contributing more to the constraints we see in Figure~\ref{fig:corner_emu_CMASS}. We radially split the derivatives being input into the Fisher analysis (see Equation~\ref{eq:fisher}) to have ``inner'' and ``outer'' components, and repeat the process described in Section~\ref{sec:methods_fisher} for each component. Here, we define the ``inner'' component to be the SZ profiles inward of 3 arcminutes, and the ``outer'' component to be the other half of the profile beyond 3 arcminutes. This cut was chosen to have an equal number of SO radial bins for each component, so the ``inner'' and ``outer'' parts each contain 3 radial points (see Figure~\ref{fig:obs_SO}).

\begin{figure*}
    \centering
    \includegraphics[scale=0.5]{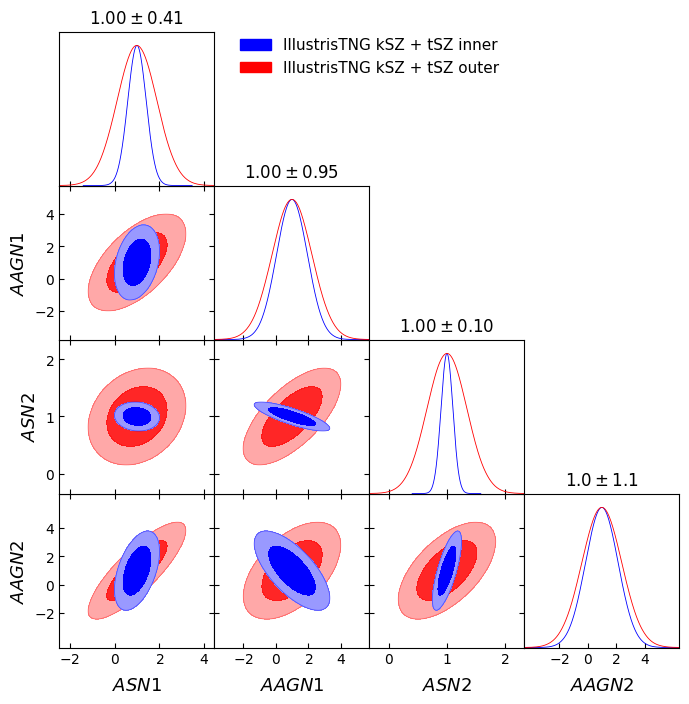}
    \includegraphics[scale=0.5]{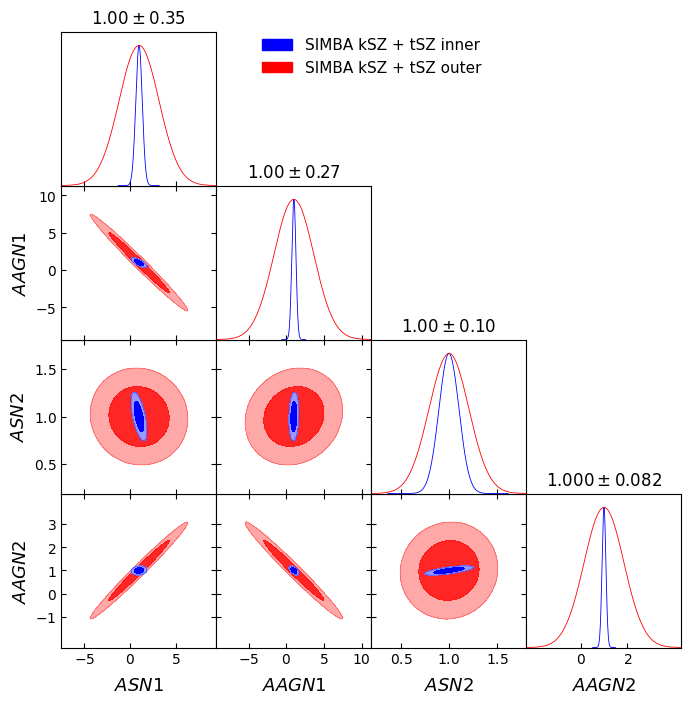}
    \caption{Corner plots showing constraints from the CMASS emulator for CAMELS-IllustrisTNG (left) and CAMELS-SIMBA (right) for the different radial components of the profiles, as discussed in Section~\ref{sec:results_constraints_split}. The blue contours show the forecasted constraints of the feedback parameter using the inner profile (inward of 3 arcminutes), and the red contours show the forecasted constraints using the outer profile. These profiles and derivatives were calculated using a Gaussian beam, and the $\pm 1\sigma$ value of the distribution for each parameter is shown above each panel on the diagonal and listed in Table~\ref{tab:emu_constraints}. It can be seen from all panels that the inner profile provides tighter contraints than the outer profile, which is expected due to the smaller error bars.}
    \label{fig:constraint_split}
\end{figure*}

The constraints from each component are shown in Figure~\ref{fig:constraint_split}, with the results from CAMELS-IllustrisTNG on the left and CAMELS-SIMBA on the right. The blue contours show the forecasted constraints from the inner profile, and the red contours show the forecasted constraints from the outer profile, both combining the information from kSZ and tSZ. The $1\sigma$ constraints for the inner profile are shown on top of each panel on the diagonal, and both inner and outer constraints, along with those from the entire profile, are shown in Table~\ref{tab:emu_constraints}. It can be seen from the Fisher analysis of this sample (with the forecasted SO error bars in particular) that the inner profile is more sensitive to the changing feedback parameters and thus has more constraining power than the outer profile. This could be due to a few reasons: first and foremost, as seen in Figures~\ref{fig:obs_SO} and~\ref{fig:parameter_derivs} and mentioned previously, our error bars increase in size as we increase in radius. With larger errors, the constraining power of the outer profiles is diminished. Second, both the SN and AGN sub-grid feedback models are centrally located, therefore the effects from these models are concentrated in the inner profile, as we can see in the derivatives of Figure~\ref{fig:parameter_derivs}. The derivatives tend to decrease as radius increases, which leads to wider constraints for the outer radial component.

This is an interesting result and highlights the complementary nature of SZ and X-ray analyses. X-ray observations can provide more information on the inner profile, since they are more sensitive to the higher density gas in these regions.
With the size of the forecasted SO error bars, it is clear that the inner profile contributes the majority of the constraining power, but we still need the combination of the two components to achieve the tightest constraints possible. As can be seen in the ``Gaussian'' columns of  Table~\ref{tab:emu_constraints}, the constraints from using the total profile are better than those from either of the individual components alone.

\subsection{Comparing CAMELS Simulations to Observations}\label{sec:results_obs}
\begin{figure*}
    \centering
    \includegraphics[scale=0.7]{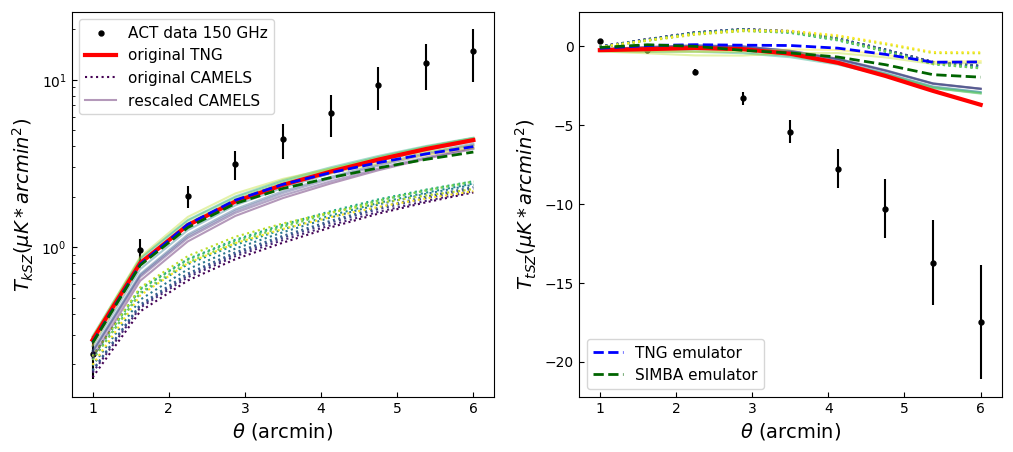}
    \caption{This figure shows simulated CAMELS kSZ (left) and tSZ (right) profiles of the CAMELS-IllustrisTNG suite varying $A_{SN2}$ as an illustration compared to observed profiles from ACT and profiles from the original IllustrisTNG simulation. The ACT data and errors at 150 GHz presented in \citet{Amodeo2021} and \citet{Schaan2021} are shown in black, along with the profiles from the original IllustrisTNG simulation shown in red. The dotted lines show the original profiles computed from the CAMELS simulations, and the solid lines show the original profiles rescaled to account for differences in simulation volumes, described in Section~\ref{sec:results_obs}. The dashed blue and green curves show the best fit profile derived by the CMASS emulator for CAMELS-IllustrisTNG and CAMELS-SIMBA, respectively, and demonstrate that the simulations under-predict the observations for any combination of parameters within the range explored by the CAMELS simulations.}
    \label{fig:obs_ACT}
\end{figure*}

We calculated simulated kSZ and tSZ profiles as described in Section~\ref{sec:methods_profiles} for profiles varying each of the feedback parameters. In Figure~\ref{fig:obs_ACT} we compare the profiles from the CAMELS-IllustrisTNG simulations varying the $A_{SN2}$ parameter (as an illustration, in dotted and solid colored lines) to ACT observations with errors at 150 GHz \citep{Amodeo2021,Schaan2021} in black and the profiles from the original IllustrisTNG (box size 100, resolution 3) simulation (see \citet{Moser2021} and Figure 6 of \citet{Amodeo2021}) in red. Additionally, using the emulator we derive the combination of feedback parameters that results in the best fit to the data, shown by blue and green dashed lines. All profiles were calculated at redshift $z=0.54$ and weighted to match the mass distribution of the CMASS sample, although the original IllustrisTNG profiles include the additional high mass halos absent in the CAMELS simulations due to the smaller simulation volume, discussed further below. All projections of the simulated profiles were convolved with the ACT beam, and the tSZ projections include a dust contribution \citep{Amodeo2021,Schaan2021}. Since the profiles are coming from the simulations, they automatically include a two-halo term contribution.

It can be seen in Figure~\ref{fig:obs_ACT} that there is a significant difference between the original IllustrisTNG100 (red) and the original CAMELS profiles (dotted). The discrepancies in these specific profiles are mainly due to the difference in simulation volumes and evolutionary histories of the halos, along with less significant ($\lesssim10\%$) effects from resolution differences (see Figure 1 of \citet{Moser2021}). Therefore, to account for the lack of massive halos due to differences in simulation volumes, we perform a rescaling of the CAMELS profiles to what we would expect them to be for the larger volume of the original IllustrisTNG100 simulation. We calculate the rescaling factor by dividing the original IllustrisTNG projected SZ profile by the fiducial CAMELS projected SZ profile, i.e., the profile for the simulation with $A_{SN2} = 1.0$. Then we multiply all of the CAMELS SZ profiles by the rescaling factor, shown by the solid colored lines in the figure, effectively increasing the amplitudes of the original CAMELS profiles to be closer to that of the original IllustrisTNG profile. We use the same rescaling process for the CAMELS-SIMBA profiles, as we expect the same kind of discrepancy due to the smaller simulation volume of the CAMELS suite.

Similarly to the results discussed in \citet{Amodeo2021}, at smaller radii ($< 2^\prime$) the simulated profiles relatively agree with the observations, but at larger radii the kSZ and tSZ profiles predicted by the simulations are significantly lower in amplitude than the measurements. We note that, as discussed earlier in Section \ref{sec:methods_fisher}, the error bars shown here have non-negligible covariances \citep{Schaan2021}, so ``chi by eye'' is not advised. As mentioned in \citet{Amodeo2021}, the discrepancy is not statistically significant for the kSZ while it is for the tSZ. The differences between the original CAMELS profiles and the observations are helped by our rescaling, but they are all still under-predicting the measurements. These simulations were not necessarily calibrated to match these specific observations. However, despite CAMELS exploring some of the feedback parameter space, differences between the simulation predictions and observations persist and are statistically significant for the tSZ. The origin of these differences remains an open question.

Lastly, we combine the individual emulators for each parameter being varied in the CAMELS suite into one emulator to attempt to find the combination of parameters resulting in the best fit CAMELS simulated profile to the ACT data, shown by the blue and green dashed lines of Figure~\ref{fig:obs_ACT}. We tune the parameters based on their known effects on the profiles (shown explicitly in Figure~\ref{fig:parameter_derivs}) and calculate the $\chi^2$ of the emulated profile to find the best combination. The purpose of this exercise is to qualitatively show how we can adjust the profiles with the emulator rather than be a quantitative study because there are caveats that make a full quantitative analysis not viable. First, the emulator is built on the \textit{1P} simulation profiles, meaning that it is only trained on profiles varying one of the parameters at a time. As such, it cannot emulate non-linear relationships between the parameters which could further change the shapes of the profiles. Second, the emulator can only interpolate the profiles it is given, so it is difficult to drastically change the shapes of the profiles, particularly the outer profiles for which the simulations significantly under-predict the observations. Lastly, the errors on the ACT measurements are roughly an order of magnitude larger than the forecasted SO errors, where we were obtaining constraints ranging from 10s of percent to 6 percent. Thus, performing likelihood-fitting analyses would provide non-informative constraints on the sub-grid parameters. 

With the above caveats in mind, the best attempts for a fit to the observations can be seen in the Figure~\ref{fig:obs_ACT} as dashed lines. For the kSZ profiles, the simulations are all essentially consistent, with the CAMELS-IllustrisTNG profile providing a slightly better fit than the original IllustrisTNG and CAMELS-SIMBA. The $\chi^2$ values for the original IllustrisTNG, CAMELS-IllustrisTNG, and CAMELS-SIMBA with respect to the observed kSZ profile are 11.37, 11.23, and 11.98, respectively. The parameter combinations for each of these profiles are $[A_{SN1}, A_{AGN1}, A_{SN2}, A_{AGN2}] = [0.84, 1.13, 1.74, 1.02]$ for CAMELS-IllustrisTNG and $[0.69, 0.25, 2.00, 0.50]$ for CAMELS-SIMBA. For the tSZ, it can be seen that none of the simulations provides a good fit to the observed data beyond 2 arcminutes. The $\chi^2$ values for the original IllustrisTNG, CAMELS-IllustrisTNG, and CAMELS-SIMBA are 324.21, 248.39, and 202.99, respectively. Similarly, the parameter combinations for these best fits are $[0.25, 1.00, 1.00, 1.00]$ for CAMELS-IllustrisTNG and $[1.00, 1.00, 1.00, 1.74]$ for CAMELS-SIMBA.
The best fit profile for CAMELS-SIMBA used the fiducial amplitudes for all of the parameters except for $A_{AGN2}$ which had the value of 1.74, close to the maximum of 2.00 explored in CAMELS. The tSZ $\chi^2$s indicate that with the variation of the feedback parameters allowed by the CAMELS suite, we are able to produce a better fit to the data than the original IllustrisTNG simulation with set values. However, these values for the tSZ profiles are high, even for the best fit of CAMELS-SIMBA. We emphasize that even with the most extreme amplitudes being varied by CAMELS, including the strong feedback model in SIMBA, we were unable to match the tSZ observations.

\section{Discussion and Conclusions}\label{sec:conclusions}
In this study we aim to forecast constraints of the thermodynamic feedback processes occurring within the CGM with simulated SZ profiles. We calculate density and pressure profiles from the CAMELS suite of simulations that varies four different feedback parameters, and then project the profiles into observable kSZ and tSZ profiles, shown by Figure~\ref{fig:profiles_3d}. We derive a general emulator for CAMELS profiles that can produce a mean/median density and pressure profile for any given redshift, mass, and feedback parameter to an accuracy of $\lesssim 10\%$, shown by Figure~\ref{fig:emu_accuracy}. This level of accuracy is better than previous parametric fitting attempts in lower dimensional spaces, for example fitting formulas used for intracluster medium profiles \citep[e.g.,][]{Nagai2007,Battaglia2012b,Battaglia2016}.

As an example for a specific observed sample, we derive an emulator for a CMASS-like sample with a DESI-like number of galaxies as observed by ACT and the future SO experiment. We show the simulated tSZ and kSZ profiles along with forecasted SO error bars in Figure~\ref{fig:obs_SO}, highlighting that with these observations we would be able to distinguish among all of the different astrophysical models being varied by CAMELS. We use the emulator to calculate derivatives of the profiles as a function of feedback parameter, and combine the derivatives into a Fisher analysis to quantify the constraints we are able to achieve, given certain error bars. We also show how the constraints vary with the convolution of different beams, testing the ACT beam against a more ideal Gaussian beam. In Figure~\ref{fig:corner_emu_CMASS} we show the forecasted constraints from the combination of tSZ and kSZ profiles, and find that for both of the beams we are able to achieve constraints of $<10\%$ on certain parameters. We list the constraints for each beam and suite in Table~\ref{tab:emu_constraints}, and find that the constraints are indeed tighter when using a Gaussian beam compared to the ACT beam. 

We also explore the different radial components of the profiles contributing to the constraints by performing separate Fisher analyses for the inner and outer profiles, shown by Figure~\ref{fig:constraint_split}. We find that for this sample and errors in particular, a larger percentage of the constraint is coming from the inner profile (inward of 3 arcminutes) compared to the outer profile. This is due to the error bars increasing in size as we increase in radius and/or the fact that all of the feedback models are centrally located, so the effects are concentrated in the inner profile. However, we do still need the combination of both inner and outer components to result in tighter constraints than either component alone (compare the ``Inner'', ``Outer'', and ``Total'' columns of Table~\ref{tab:emu_constraints}). 

We show the comparison of the simulated CAMELS profiles to the observed profiles from ACT in Figure~\ref{fig:obs_ACT}, and find that the simulated CAMELS profiles under-predict the observations similarly to the original IllustrisTNG simulation. Additionally, we determine the combination of feedback parameters that results in the best fit of the CAMELS simulations to the observed profiles, and find that for kSZ we are able to find a slightly better fit from CAMELS-IllustrisTNG than the original IllustrisTNG simulation. For the tSZ profiles, we are able to find a better fit to the observations from CAMELS-SIMBA compared to the original IllustrisTNG, but in general none of the simulated profiles are able to provide good fits. To match the current kSZ and tSZ observations, the simulations need to be able to predict both higher gas density and pressure in halos.

The \textit{1P} CAMELS simulation set offers a wide range of parameter variation, so the inability of these simulations to match the observations is important to investigate further. As previously mentioned, a few limitations to this study include the small simulation volumes resulting in few high mass objects and the emulator being built on the \textit{1P} set only sensitive to changes in one parameter at a time. One possible future experiment would be expanding these methods to construct an emulator on simulations that vary more than one parameter at a time, such as the CAMELS Latin-hypercube set. This \textit{LH} set contains 1,000 simulations varying all 6 parameters (4 astrophysical and 2 cosmological) and could potentially capture differences in the profiles due to changes in multiple parameters at once.

In future studies, the constraints on feedback processes in the CGM and IGM would improve with higher quality observations (including smaller error bars), and the combination of different observations such as SZ and X-ray to focus on the different components of the radial profiles. X-ray observations will be useful in constraining the inner regions, since they are more sensitive to the higher density gas and they have higher angular resolution than SZ observations. Current datasets like the eROSITA Final Equatorial-Depth Survey \citep[eFEDS;][]{Predehl2021,Brunner2021} have potential to further constrain these astrophysical models. 
Additionally, combining density and pressure information with other kinds of tracers such as metallicity profiles from line measurements could also potentially help in assessing the validity of different feedback models. Finally, expanding these methods to use more simulations that vary multiple models at once and including more simulation frameworks in addition to IllustrisTNG and SIMBA would be beneficial in exploring many different physical feedback model implementations. Here we explored the variation of 4 specific astrophysical parameters from IllustrisTNG and SIMBA, but in general, such simulations have many more models and parameters with varying levels of constraints. It is not obvious that we would draw similar conclusions in our comparisons to current SZ observations if we were to vary an even higher number of parameters or perhaps a different sub-grid parameter we have not considered here. Going forward, the key to further constraining models in simulations is to use all possible information, including SZ, X-rays, galaxy surveys, line measurements, etc. to extract the maximum amount of information from the sources we have available.

\section*{Acknowledgements}
We thank Shy Genel, Dylan Nelson, Rachel Somerville, Emmanuel Schaan, and Simone Ferraro for their helpful comments. EM and NB are supported by NSF grant AST-1910021 and NB acknowledges support from NASA grants 21-ADAP21-0114 and 21-ATP21-0129. DAA was supported in part by NSF grants AST-2009687 and AST-2108944.
The Flatiron Institute is supported by the Simons Foundation. 

\section*{Data Availability}
The CAMELS simulations and thermodynamic profiles presented in this work are publicly available \citep{Villaescusa-Navarro2022}. Further details and instructions on data usage can be found at \url{https://camels.readthedocs.io} and \url{https://www.camel-simulations.org}.

\bibliographystyle{aasjournal}
\bibliography{cits.bib}

\end{document}